\documentclass[twocolumn,a4paper,reprint,amssymb,aps,prd,groupedaddress,superscriptaddress,amsmath,nobibnotes,showpacs,nofootinbib]{revtex4-2}
\usepackage{graphicx,bm,color,psfrag}
\usepackage{amsfonts}
\usepackage{lipsum}
\usepackage{enumitem}

\usepackage{mathtools}

\usepackage[colorlinks = true,
            linkcolor = blue,
            urlcolor  = blue,
            citecolor = blue,
            anchorcolor = blue]{hyperref}
\usepackage{soul}
\usepackage{xcolor}
\usepackage{threeparttable,booktabs}

\newcommand{\be}{\begin{equation}}
\newcommand{\ee}{\end{equation}}

\begin{document}

\title{Weak field and slow motion limits in energy-momentum powered gravity}

\author{\"{O}zg\"{u}r Akarsu}
\email{akarsuo@itu.edu.tr}
\affiliation{Department of Physics, Istanbul Technical University, Maslak 34469 Istanbul, Turkey}

\author{A. Kaz{\i}m \c{C}aml{\i}bel}
\email{camlibel@tau.edu.tr}
\affiliation{Department of Electrical and Electronics Engineering Turkish-German University,  Beykoz 34820 Istanbul, Turkey}

\author{Nihan Kat{\i}rc{\i}}
\email{nkatirci@dogus.edu.tr}
\affiliation{Department of Electrical and Electronics Engineering Do\u gu\c s University, \"Umraniye 34775 Istanbul, Turkey}

\author{\.{I}brahim Semiz}
\email{semizibr@boun.edu.tr}
\affiliation{Department of Physics, Bo\u{g}azi\c{c}i University, Bebek 34342 Istanbul, Turkey}

\author{N. Merve Uzun}
\email{nebiye.uzun@boun.edu.tr}
\affiliation{Department of Physics, Bo\u{g}azi\c{c}i University, Bebek 34342 Istanbul, Turkey}

\begin{abstract}
We explore the weak field and slow motion limits, Newtonian and Post-Newtonian limits, of the energy-momentum powered gravity (EMPG), viz., the energy-momentum squared gravity (EMSG) of the form $f(T_{\mu\nu}T^{\mu\nu})=\alpha (T_{\mu\nu}T^{\mu\nu})^{\eta}$ with $\alpha$ and $\eta$ being constants. We have shown that EMPG with $\eta\geq0$ and general relativity (GR) are not distinguishable by local tests, say, the Solar System tests; as they lead to the same gravitational potential form, PPN parameters, and geodesics for the test particles. However, within the EMPG framework, $M_{\rm ast}$, the mass of an astrophysical object inferred from astronomical observations such as planetary orbits and deflection of light, corresponds to the effective mass $M_{\rm eff}(\alpha,\eta,M)=M+M_{\rm empg}(\alpha,\eta,M)$, $M$ being the actual physical mass and $M_{\rm empg}$ being the modification due to EMPG. Accordingly, while in GR we simply have the relation $M_{\rm ast}=M$, in EMPG we have $M_{\rm ast}=M+M_{\rm empg}$. Within the framework of EMPG, if there is information about the values of $\{\alpha,\eta\}$ pair or $M$ from other independent phenomena (from cosmological observations, structure of the astrophysical object, etc.), then in principle it is possible to infer not only $M_{\rm ast}$ alone from astronomical observations, but $M$ and $M_{\rm empg}$ separately. For a proper analysis within EMPG framework, it is necessary to describe the slow motion condition (also related to the Newtonian limit approximation) by $|p_{\rm eff}/\rho_{\rm eff}|\ll1$ (where $p_{\rm eff}=p+p_{\rm empg}$ and $\rho_{\rm eff}=\rho+\rho_{\rm empg}$), whereas this condition leads to $|p/\rho|\ll1$ in GR.

\end{abstract}

\maketitle
\section{Introduction}
As is well known, the current standard, or ``concordance'' model of cosmology is the $\Lambda$CDM model. In this model, the structure and dynamics of the universe is governed by general relativity (GR) with the addition of the cosmological constant $\Lambda$, where the latter is used to explain the late-time acceleration of the universe discovered around the turn of the millennium~\cite{Riess:1998cb,eBOSS:2020yzd,duMasdesBourboux:2020pck,Ade:2015xua,Aghanim:2018eyx,SCP:1998vns,DES:2021wwk}; and the cold dark matter (CDM) component, originally postulated to explain the unexpected large velocities of the galaxies
within the Coma cluster, plays also a role in the early evolution of the universe, and structure formation. The $\Lambda$CDM model, which for many years has been considered quite successful in explaining a wide range of astrophysical and cosmological observations, has recently begun to suffer from tensions of various degrees of significance with newly acquired data, with increasing precision and variety of cosmological observations~\cite{DiValentino:2020vhf,DiValentino:2020zio,DiValentino:2020vvd,DiValentino:2020srs,DiValentino:2021izs,Perivolaropoulos:2021jda,Abdalla:2022yfr}. These tensions likely point to the need for new physics, either on the side of the contents of the universe, or on the side of gravitation, i.e., how the contents influence the geometry of the spacetime. This latter option would lead to alternative/modified gravity theories as replacement of GR~\cite{Copeland:2006wr,DeFelice:2010aj,Nojiri:2010wj,Clifton:2011jh,Capozziello:2011et,Nojiri:2017ncd}.

However, precise Solar System observations are totally compatible with GR so far. In fact, the vast majority of the phenomena can be explained, albeit with limited accuracy, even by Newton's theory of gravity. Adopting an alternative theory should not undermine these achievements; yet most of them do. For example, in possibly the simplest extension, the Brans-Dicke theory, the model parameter was constrained as $\omega \gtrsim 40000$ at the 2$\sigma$ confidence level from telemetry observations of the  Cassini spacecraft~\cite{Bertotti:2003rm}, whereas the initial formulation of the theory had anticipated $\omega \sim \mathcal{O}(1)$ ($\omega \longrightarrow \infty$ is the GR limit). In the most common generalization of GR, $f(\mathcal{R})$ theories, the scalar curvature $\mathcal{R}$ satisfies a second order differential equation  which induces non-Newtonian, Yukawa-type exponential potentials in the linearized level~\cite{Chiba:2003ir}.~\footnote{The Schwarzchild metric is not the unique vacuum solution in this case, since Birkoff's theorem does not hold (see~\cite{Olmo:2019flu} for details).} For instance, a particular choice, $f(\mathcal{R})=\mathcal{R}-\mu^4/\mathcal{R}$ was ruled out by data from Solar System tests. It has also been shown that $f(\mathcal{R})$ gravity is strongly constrained by local observations~\cite{olmo05,olmo055,olmo07} in the laboratory and solar scales and can only evade these tight gravitational tests through some screening mechanisms~\cite{brax10,brax12,brax122}.

In the settings close to the Newtonian limit, the parametrized post-Newtonian (PPN) formalism  describes the metric of any gravitational theory with standard potentials and ten parameters (only $\gamma = \beta = 1$ being nonzero in GR) whose values can characterize the theory under consideration. The compatibility of the Solar System observations referred to in the last paragraph is reflected in  the calculated parameters $\gamma$ and $\beta$ being very close to 1 and others very close to zero. For the current limits on all PPN parameters, we refer the reader to Table~4 in~\cite{Will:2014kxa}. The strongest bound for $\gamma$ (which controls the deflection of light) is determined by the Shapiro time delay measurement of the Cassini--Huygens tracking experiment, giving $\gamma=1+(2.1\pm 2.3)\times 10^{-5}$~\cite{Bertotti:2003rm,Shapiro:2004zz}. Adopting the Cassini bound on $\gamma$, the latest constraints on $\beta$ (which controls the perihelion precession of Mercury) are $\beta =1+(-4.1\pm 7.8)\times 10^{-5}$ from helioseismology data~\cite{Mecheri2004NewVO, Antia:2007hm} and $\beta=1+(1.2\pm 1.1)\times 10^{-4}$ from the lunar laser ranging experiment~\cite{Williams:2004qba, Williams:2005rv}. Obviously, any theory of gravitation should also satisfy these bounds on the values of PPN parameters to be considered viable, regardless of how well it explains cosmological observations. In fact, there is at least one well-known modified theory of gravity that alters cosmological dynamics while leaving the predictions at solar scales unchanged compared to those of GR; namely the teleparallel equivalent of general relativity (TEGR), which gives $\gamma=\beta=1$ as in GR~\cite{Bahamonde:2021gfp}; however, the concept of mass for the central object seems ambiguous in this theory~\cite{DeBenedictis:2016aze,Olmo:2019flu}.

In the current paper, we will examine a specific version of energy-momentum squared gravity (EMSG)~\cite{Katirci:2014sti,Akarsu:2017ohj,Board:2017ign}, known as energy-momentum powered gravity (EMPG)~\cite{Akarsu:2017ohj,Board:2017ign}, to show that it leaves the PPN parameters as in GR and to clarify the mathematics and interpretation of weak field and slow motion limits within the framework of EMSG---a framework closely related to modified gravity theories such as $\mathcal{R}+f(\mathcal{L}_{\rm m})$ \cite{Harko:2010mv} and $\mathcal{R}+f(T)$ \cite{Harko:2011kv}, as all which are identical in that the source is minimally coupled to the curvature. The EMSG theory~\cite{Katirci:2014sti,Akarsu:2017ohj,Board:2017ign}, generalizes the matter sector of GR with the arbitrary function of the Lorentz scalar $T_{\mu\nu} T^{\mu\nu}$, viz., $f(T_{\mu\nu} T^{\mu\nu})$ included in the usual Einstein-Hilbert action---here, $T_{\mu\nu}$ is the energy-momentum tensor (EMT), $T$ above is its trace, $\mathcal{L}_{\rm m}$ is the matter Lagrangian density and the ``squared'' terminology arises from the self contraction of EMT. The EMSG theory has been studied so far mainly in cosmological and partially in astrophysical contexts~\cite{Katirci:2014sti,Akarsu:2017ohj,Board:2017ign,Roshan:2016mbt,Akarsu:2018zxl,Nari:2018aqs,Akarsu:2018aro,Akarsu:2019ygx,Faria:2019ejh,Bahamonde:2019urw,Chen:2019dip,Barbar:2019rfn,Kazemi:2020hep,Singh:2020bdv,Nazari:2020gnu,Rudra:2020rhs,Akarsu:2020vii,Chen:2021cts,Nazari:2022xhv,Nazari:2022fbn,Tangphati:2022acb,Khodadi:2022xtl,Acquaviva:2022bju,Khodadi:2022zyz,Akarsu:2023nyl}.

Some interesting features of the EMSG theory studied in the framework of various models are the nonconservation of EMT~\cite{Akarsu:2017ohj,Board:2017ign}; the possibility of driving late time acceleration from the usual cosmological sources without a cosmological constant $\Lambda$~\cite{Akarsu:2017ohj}; the screening of $\Lambda$ in the past by the new contributions of dust in the Friedmann equation~\cite{Akarsu:2019ygx}; the effective source that yields constant inertial mass density arises in the energy-momentum log gravity (EMLG)~\cite{Akarsu:2019ygx}; the screening of the shear scalar (viz., the contribution of the expansion anisotropy to the average expansion rate of the universe) via the new contribution of dust in the Friedmann equation in quadratic EMSG~\cite{Akarsu:2020vii}---this model can lead to exactly the same Friedmann equation of the standard $\Lambda$CDM model even with anisotropic expansion; the altered past or far future of the Universe~\cite{Roshan:2016mbt,Akarsu:2018zxl,Acquaviva:2022bju}; etc. The particular model of EMSG we consider in this work is the EMPG~\cite{Akarsu:2017ohj,Board:2017ign}, a specific model of EMSG with the choice of $f(T_{\mu\nu}T^{\mu\nu})= \alpha (T_{\mu\nu}T^{\mu\nu})^{\eta}$ where $\alpha$ is the coupling parameter and $\eta$ is a constant (it controls whether the EMPG modification will be more effective at large energy density scales/in the early universe or low energy density scales/in the late universe); and we will study EMPG in the weak field, slow motion (Newtonian and post-Newtonian) limit. 
 Using the effective EMT interpretation, we show that there is a direct analogy between the effective matter variables of EMSG and the standard matter variables of GR which provides us with the most general slow-motion condition of the model. We illustrate that the Poisson's equation remains the same as in GR, except that the standard energy density $\rho$ is replaced by the effective energy density $\rho_{\rm eff}$. Recall that in astronomy we do not usually measure/observe the mass of an astronomical object directly, but actually find it by Keplerian methods, namely, inferring from nearby orbits, e.g., from the measurements/observations of the planet's/satellite's orbital semi-major axis and period revolving about it. As there is no modification in the curvature sector of the theory, viz. spacetime is still governed by GR, albeit, with the effective EMT, that EMT can be solely inferred via astronomical methods. Hence, the mass (energy) density that we infer with such methods is the effective mass (energy) density, i.e., the conventional Newtonian potential is determined by $M_{\rm eff}$ instead of $M$. An important corollary is that one component of the Newtonian limit approximation, the slow motion condition, $|p/\rho|\ll1$, will have to be replaced by $|p_{\rm eff}/\rho_{\rm eff}|\ll1$ if one analyses astrophysical objects. Note that the former may not necessarily be satisfied in cases where the latter is.

The paper is structured as follows: In Section~\ref{sec:EMPG}, we present the detailed framework of EMPG;  in Section~\ref{sec:wf}, we derive the equations of motion (EoM) in the linearized theory of EMPG, discuss the slow motion condition, Newtonian and post-Newtonian limits of the model; in Section~\ref{sec:exterior}, we confirm our results from the Schwarzschild exterior solution; in Section~\ref{discussion}, we presented an assessment on constraining $\alpha$ and $\eta$ from the slow motion condition; and in Section~\ref{conclusion}, we draw final conclusions from our results.

\section{energy-momentum powered gravity }
\label{sec:EMPG}
The action for EMSG is
\begin{align}
S=\int {\rm d}^4x\,\sqrt{-g}\left[\frac{1}{2\kappa} \mathcal{R}+f(T_{\mu\nu}T^{\mu\nu})+\mathcal{L}_{\rm m}\right],
\label{action}
\end{align}
where $\kappa=8 \pi G$ ($G$ is Newton's constant), $g$ is the determinant of the metric tensor $g_{\mu\nu}$, $\mathcal{R}$ is its Ricci scalar, $\mathcal{L}_{\rm m}$ is the Lagrangian density describing the matter source and $T_{\mu\nu}$ is its EMT; and $f$ is an arbitrary function of the Lorentz scalar $T_{\mu\nu}T^{\mu\nu}$~\cite{Katirci:2014sti,Akarsu:2017ohj,Board:2017ign}. Here and throughout the paper, we work in units such that the speed of light, $c$, equals unity. The variation of this action with respect to the inverse metric $g^{\mu\nu}$ is
\begin{equation}
 \begin{aligned}   \label{variation}
  \delta S=\int\, {\rm d}^4 x \bigg[&\frac{1}{2 \kappa}\frac{\delta (\sqrt{-g}\mathcal{R})}{\delta g^{\mu\nu}}+\frac{\delta(\sqrt{-g}\mathcal{L}_{\rm m})}{\delta g^{\mu\nu}}      \\
 &+f(T_{\sigma\epsilon}T^{\sigma\epsilon}) \: \frac{\delta(\sqrt{-g})}{\delta g^{\mu\nu}} \\
 &+\frac{\partial f}{\partial(T_{\lambda\xi}T^{\lambda\xi})}\frac{\delta(T_{\sigma\epsilon}T^{\sigma\epsilon})}{\delta g^{\mu\nu}}\sqrt{-g}\bigg] \delta g^{\mu\nu},  
\end{aligned}
\end{equation}
and, as usual, the definition of the standard EMT in terms of the matter Lagrangian density $\mathcal{L}_{\rm m}$ reads
  \begin{equation}
  \label{tmunudef}
 T_{\mu\nu}=-\frac{2}{\sqrt{-g}}\frac{\delta(\sqrt{-g}\mathcal{L}_{\rm m})}{\delta g^{\mu\nu}}=g_{\mu\nu}\mathcal{L}_{\rm m}-2\frac{\partial \mathcal{L}_{\rm m}}{\partial g^{\mu\nu}},
 \end{equation}
for which we suppose $\mathcal{L}_{\rm m}$ depends only on the metric tensor components, and not on its derivatives; because the matter fields usually couple only to the metric and not to its derivatives---this is the case for the Maxwell field and gauge fields in general, as well as for scalar fields.
 
Hence, the field equations---or EoM---for EMSG are
\begin{equation}
G_{\mu\nu} =\kappa  \left( T_{\mu\nu}+f g_{\mu\nu}-2\frac{\partial f}{\partial(T_{\sigma\epsilon}T^{\sigma\epsilon})}\theta_{\mu\nu}\right),
\label{modfieldeq}
\end{equation}
where $G_{\mu\nu}= \mathcal{R}_{\mu\nu}-\frac{1}{2}\mathcal{R}g_{\mu\nu}$ is the Einstein tensor,
\begin{equation}
\begin{aligned}
\theta_{\mu\nu}&=-2\mathcal{L}_{\rm m}\left(T_{\mu\nu}-\frac{1}{2}g_{\mu\nu}T\right)-T T_{\mu\nu}\\
&\quad\,+2T_{\mu}^{\lambda}T_{\nu\lambda}-4T^{\sigma\epsilon}\frac{\partial^2 \mathcal{L}_{\rm m}}{\partial g^{\mu\nu} \partial g^{\sigma\epsilon}},
\label{theta}
\end{aligned}
\end{equation}
which is a new tensor defined as $\theta_{\mu\nu}\equiv \delta (T_{\sigma\epsilon}T^{\sigma\epsilon})/\delta g^{\mu\nu}$, and $T=g_{\mu\nu}T^{\mu\nu}$ is the trace of the EMT.

We proceed with a particular form of EMSG, the powered form of the Lorentz scalar $T_{\mu\nu}T^{\mu\nu}$ in the action~\eqref{action}, known as EMPG, described by 
\begin{equation}
\label{eqn:powerassumption}
f(T_{\mu\nu}T^{\mu\nu})= \alpha (T_{\mu\nu}T^{\mu\nu})^\eta,
\end{equation}
where $\eta$ is the dimensionless constant parameter that determines the power of the energy-momentum squared term and $\alpha$ is the constant parameter that measures the gravitational coupling strength of the EMPG modification to GR (note that $\alpha$'s dimension depends on $\eta$)~\cite{Akarsu:2017ohj,Board:2017ign}. Accordingly, from  Eqs.~\eqref{modfieldeq}, the field equations for EMPG are
\begin{equation}
G_{\mu\nu}=\kappa \left[ T_{\mu\nu}+\alpha (T_{\sigma\epsilon}T^{\sigma\epsilon})^{\eta}\left(g_{\mu\nu}-2 \eta \frac{\theta_{\mu\nu}}{T_{\lambda\xi}T^{\lambda\xi}}\right)\right],
\label{fieldeq}
\end{equation}
in which the new terms arising from the EMPG modification,
\begin{align} 
\label{emtEMPG}
T_{\mu\nu}^{\rm empg}\equiv&\,\alpha (T_{\sigma\epsilon}T^{\sigma\epsilon})^{\eta}\bigg(g_{\mu\nu}-2 \eta \frac{\theta_{\mu\nu}}{T_{\lambda\xi}T^{\lambda\xi}}\bigg),
\end{align}
vanish in the $\alpha \rightarrow 0$ limit in an obvious way, independent of $\eta$. We call all the matter-related terms on the right-hand side of Eq.~\eqref{fieldeq} \textit{effective} EMT, $T_{\mu\nu}^{\rm eff}$. It follows that
\begin{align}
\label{emteff}
T_{\mu\nu}^{\rm eff} \equiv &\,T_{\mu\nu}+T_{\mu\nu}^{\rm empg},
\end{align}
and, as the $G_{\mu\nu}$ is fully determined by this term, according to the EoM~\eqref{fieldeq} and definitions~(\ref{emtEMPG}) and~(\ref{emteff})
\begin{equation} \label{FEwithG}
G_{\mu\nu}=8\pi G \,T_{\mu\nu}^{\rm eff}.
\end{equation}
This means that the spacetime geometry side of the field equations is described exactly as in GR, while the material side is described by the effective EMT composed of the standard and EMPG modification parts. From here on, we will call the EMT defined by Eq.~(\ref{tmunudef}) the GR-standard EMT. As the modifications appear only in the material side of the field equations in contrast to the case with $f(\mathcal{R})$ and Brans-Dicke type theories, it is sufficient to determine the components of the effective EMT in our model, more generally in any particular model of EMSG.

The twice-contracted Bianchi identity, $\nabla^{\mu}G_{\mu\nu}=0$, implies from Eq.~\eqref{FEwithG} that the divergence of the effective EMT always vanishes;
\begin{align}
\label{nonconservedenergy}
\nabla^{\mu}T_{\mu\nu}^{\rm eff}=0.
\end{align}
Note however from Eq.~\eqref{emteff} that the covariant derivatives of $T_{\mu\nu}$ and $T_{\mu\nu}^{\rm empg}$ do not necessarily vanish, namely, the matter EMT is not necessarily conserved ($\nabla^{\mu}T_{\mu\nu} = 0$) in EMPG. As matter source, let us take the usual perfect fluid, which is often used in settings ranging from compact objects to cosmology;
\begin{align}
\label{em}
T_{\mu\nu}=(\rho+p)u_{\mu}u_{\nu}+p g_{\mu\nu},
\end{align} 
where $\rho>0$ and $p$ are, respectively, the fluid's energy density and thermodynamic pressure measured by an observer moving with the fluid, $u^{\mu}=\frac{{\rm d} x^{\mu}}{{\rm d} \tau}$ is the fluid's four-velocity satisfying the condition $u_{\mu}u^{\mu}=-1$. Here $\tau$ is the proper time along the world lines, $t=x^0$ is the coordinate time, and $x^{\mu}=x^{\mu}(\tau)$ are the local spacetime coordinates with $\mu=0,1,2,3$. 

As the definition of the matter Lagrangian density that gives rise to the EMT of a perfect fluid is not unique, one could choose either $\mathcal{L}_{\rm m}=p$ or $\mathcal{L}_{\rm m}=-\rho$, which results in the same EMT, viz., the $T^{\mu\nu}$ that describes perfect fluid matter distributions as given in Eq.~\eqref{em}. In this study, following the literature to date on EMSG and on similar theories (see~\cite{Harko:2011kv,Haghani:2013oma,Odintsov:2013iba,Asimakis:2022jel}), we consider $\mathcal{L}_{\rm m}=p$. Hence, the Lorentz scalar $T_{\sigma\epsilon}T^{\sigma\epsilon}$ and the tensor $\theta_{\mu\nu}$ read\footnote{In accordance with the convention in the literature on EMSG so far, we omitted the $\frac{\partial^2 \mathcal{L}_{\rm m}}{\partial g^{\mu \nu} \partial g^{\sigma\epsilon}}$ term in the expression of the new tensor $\theta_{\mu\nu}$ given in Eq.~\eqref{theta}, and for the relevant discussion, we refer the reader to the recent work Ref.~\cite{Akarsu:2023lre}, appeared on arXiv during the review process of the current paper.}
\begin{align}
T_{\sigma\epsilon}T^{\sigma\epsilon}&= \rho^2+3p^2, \label{perfT2} \\
\theta_{\mu\nu}&=-  (\rho+p) (\rho+3p) u_{\mu} u_{\nu}. \label{perftheta}
\end{align}

Substituting Eq.~\eqref{em} along with Eqs.~\eqref{perfT2} and~\eqref{perftheta} into the effective EMT defined in Eq.~\eqref{emteff}, we obtain
\begin{equation}
\begin{aligned} 
\label{rhopeff}
T_{\mu\nu}^{\rm eff}=& (\rho+p)u_{\mu} u_{\nu}+p g_{\mu\nu}\\
&+\alpha (\rho^{2}+3p^2)^\eta \bigg[g_{\mu\nu} 
+2 \eta \bigg(1+\frac{4 \rho p}{\rho^2+3p^2}\bigg)u_{\mu} u_{\nu} \bigg].
\end{aligned}
\end{equation}
It can be seen that the terms in the above equation are multiplied by either $u_{\mu} u_{\nu}$ or $g_{\mu\nu}$ which makes it possible  to  write it in the standard perfect fluid EMT form
\begin{align} 
\label{effemt}
T_{\mu\nu}^{\rm eff}=(\rho_{\rm eff}+p_{\rm eff})u_{\mu}u_{\nu}+p_{\rm eff} g_{\mu\nu},
\end{align}
where $\rho_{\rm eff}$ and $p_{\rm eff}$ are, respectively, the effective energy density and effective pressure identified as
\begin{equation}
\begin{aligned}
\rho_{\rm eff}&=\rho-\alpha (\rho^2+3p^2)^{\eta}\bigg[1-2\eta \bigg(1+\frac{4 \rho p}{\rho^2+3p^2}\bigg)\bigg], \label{eq:effdens} \\
p_{\rm eff}&=p+ \alpha (\rho^2+3p^2)^\eta. 
\end{aligned}
\end{equation}
We note that the form Eq.~\eqref{effemt} for the effective EMT\footnote{From Eqs.~\eqref{nonconservedenergy} and~\eqref{effemt}, we have 
$\nabla_{\mu}(\rho_{\rm eff} u^{\mu})=-p_{\rm eff} \, \nabla_{\mu} u^{\mu}$ which along with~\eqref{eq:effdens} gives 
$\nabla_{\mu}\{[\rho-\alpha (1-2\eta) \rho^{2\eta}] u^{\mu}\}=-\alpha\rho^{2\eta}\,\nabla_{\mu} u^{\mu}$
for dust ($p=0$), then we see that the matter--current conservation equation in this model would lead to a matter creation/annihilation on cosmological scales in an expanding $(H>0)$ universe since $\nabla_{\mu}u^{\mu}=3H$ where $H$ is the Hubble parameter and $p_{\rm eff}$ is not necessarily zero, for dust as well. Also, $\nabla_{\mu}u^{\mu} \neq 0$ for a galactic system due to the stellar velocity and the galactic potential, hence, the matter creation/annihilation occurs within galaxies as well. The diversity in rotation curves of spiral galaxies and the number of observed satellites of the
Milky Way remain as unresolved issues in the CDM paradigm of the standard GR~\cite{Adhikari:2022sbh}. As a possible solution, Refs.~\cite{Spergel:1999mh,Pellicer:2011mw,Kamada:2016euw} suggest a mechanism in which DM particles exchange energy by colliding with one another giving rise to deviations in the inner halo structure. Similarly, our model that anticipates the creation/annihilation of DM can be tested by its effects on the galactic and subgalactic scales.
} together with the field equation~\eqref{FEwithG} allows us to make an analogy between the effective matter variables ($\rho_{\rm eff},p_{\rm eff}$) of EMPG and the standard matter variables ($\rho,p$) of GR, that is, $\rho_{\rm eff} \longleftrightarrow \rho$ and $p_{\rm eff} \longleftrightarrow p$. And, it can be seen directly from Eq.~\eqref{eq:effdens} that the power $\eta$ is necessarily nonnegative for $\rho_{\rm eff}$ and $p_{\rm eff}$ not to be divergent as $\rho \rightarrow 0$, when the equation of state (EoS) is assumed to be barotropic, i.e., $p=p(\rho)$; satisfying the extra condition $p(\rho=0)=0$. Consequently, for a viable effective source described by the standard perfect fluid EMT we must have $\eta \geq 0$.

\section{The Weak-Field Limit}  \label{sec:wf}
Let us recall that in any metric theory of gravitation, gravity is explained as a manifestation of the curvature of spacetime. Our aim in this work is to investigate the weak-field behaviour of EMPG, that is, behaviour close to the case of no gravity---hence no curvature, i.e., flat spacetime (Note that we did not make any assumptions on the spacetime metric $g_{\mu\nu}$ so far). Therefore, we suppose spacetime to be almost flat, so that the metric $g_{\mu\nu}$ of the curved spacetime can be split as $g_{\mu\nu}=\eta_{\mu\nu}+h_{\mu\nu}$, where $\eta_{\mu\nu}$ is the Minkowski spacetime metric representing the flat spacetime and $h_{\mu\nu}$ is a small perturbation on top of it. Then, the Ricci tensor to first order in $h_{\mu\nu}$ reads
\begin{equation}
\begin{aligned}
  \mathcal{R}^{(1)}_{\mu\nu}=\frac{1}{2}(\partial^{\alpha}\partial_{\mu} h_{\nu\alpha}+ \partial^{\alpha}\partial_{\nu} h_{\mu\alpha}-\Box h_{\mu\nu}
   -\partial_{\mu}\partial_{\nu} h),
   \end{aligned}
\end{equation}
where $\Box=\eta^{\mu\nu}\partial_{\mu}\partial_{\nu}$ is the d'Alembert operator and $h$ is the trace of $h_{\mu\nu}$.
Next, we impose, without loss of generality, the Lorenz-gauge specified by $g^{\mu\nu} \Gamma^{\alpha}_{\mu\nu}=0$; for the first order perturbation, this becomes $\partial^{\beta} h_{\alpha\beta}-\frac{1}{2} \partial_{\alpha} h=0$. In this gauge, the first order Ricci tensor simplifies to
\begin{align}
  \mathcal{R}^{(1)}_{\mu\nu}=-\frac{1}{2} \Box h_{\mu\nu}.
  \end{align}
Since $G_{\mu\nu}$ is the trace-reversed tensor of $\mathcal{R}_{\mu\nu}$, we define the trace-reversed perturbation as $\bar h_{\mu\nu}=h_{\mu\nu}-\frac{1}{2}\eta_{\mu\nu}h$ so that $G^{(1)}_{\mu\nu}=-\frac{1}{2} \Box \bar h_{\mu\nu}$. Thus, the field equations~\eqref{FEwithG} for EMPG become
\begin{equation}  
\label{linfe}
    -\Box \bar h_{\mu\nu}=16\pi G\, T_{\mu\nu}^{\rm eff}
\end{equation}
 in the weak field limit. Note that this is the same as what one would obtain for GR, except for the addition of the ``eff" label, i.e., the only difference here is that in GR the cause of gravity $T_{\mu\nu}$, whereas in our model it is $T_{\mu\nu}^{\rm eff}$.

 On the other hand, the field equations can be written in an alternative and equivalent form with the Ricci tensor on the left-hand side rather than the Einstein tensor:
  \begin{align}   \label{FEwithR}
   \mathcal{R}_{\mu\nu}=8 \pi G \left( T^{\rm eff}_{\mu\nu}-\frac{1}{2} g_{\mu\nu} T^{\rm eff}\right).
\end{align}
 This form is valid for both GR and EMPG, without/with the label ``eff". Then, to first order, we obtain
 \begin{align}  \label{linfeR}
     -\Box  h_{\mu\nu}= 16 \pi G \left(T^{\rm eff}_{\mu\nu}-\frac{1}{2} g_{\mu\nu} T^{\rm eff}\right),
\end{align}
 and will employ this form in the following subsection to illustrate a point.

\subsection{Newtonian and Post-Newtonian limits}

Although conceptually very different, Newton’s law of universal gravitation is the limiting case of Einstein's field equations when gravity is everywhere weak, and all speeds are very small compared to the speed of light ($v \ll 1$). Therefore, any alternative theory of gravitation must also be studied in this limit to make contact with the observational/experimental Solar System constraints, as discussed in the Introduction.

The first of these limits, as discussed above, leads to weakly-curved spacetime deviating {only} slightly from the flat (Minkowski) spacetime metric, governed by linearized theory, whose equations are in the form of Eq.~\eqref{linfe}, alternatively, Eq.~\eqref{linfeR}, for both GR and EMPG (without and with ``eff" label, respectively). Note that we have made no assumption on the four-velocity $u^{\mu}$ associated with the fluid yet.

In an almost flat spacetime, the four velocity can be written as $u^{\mu}=\frac{1}{\sqrt{1-v^2}}(1,v^{i})$, where $v^{i}=\frac{{\rm d} x^{i}}{{\rm d} t}$ is the three-velocity field having the magnitude of $v$~\cite{Anderson:1975} with $i=1,2,3$ reserved for spatial coordinates. For the second limit, we now concentrate on the matter distributions subject to the slow motion condition defined as all speeds being very small compared to the speed of light ($v \ll 1$), which implies the following hierarchy between the components of the effective EMT
\begin{equation}
\label{velemt}
 T_{0i}^{\rm eff}/T_{00}^{\rm eff} \sim v\qquad,\qquad T_{ij}^{\rm eff}/T_{00}^{\rm eff} \sim v^2,
\end{equation}
cf. Ref.~\cite{Will:2018bme}. It is noteworthy that when the gravitational field is weak (low curvature) and the particle is nearly at rest in the chosen reference frame, then the proper time ($\tau$) clock runs at nearly the same rate as coordinate time ($t$) clock, viz., ${\rm d}t\simeq{\rm d}\tau$, which implies
 \begin{equation}
 v_i\simeq u_i\simeq v\ll1.    
 \end{equation}
  Notice that according to the hierarchy stated in Eq.~\eqref{velemt}, regarding the slow motion condition imposed in
   EMPG and also in any other gravity theory comprising matter-type modifications minimally coupled to the curvature, the proper approach is to take into account the \textit{effective} matter variables rather than the standard ones. We emphasize that in this type of theories, the slow motion condition needs to be handled carefully; for instance, in our model, even if the effective pressure is negligible, we can still have nonnegligible standard pressure or vice versa.

\subsubsection{Newtonian limit}
 
 Assuming the metric is diagonal and time-independent ($\partial_0 h_{\mu\nu}=0$, i.e., $\Box=\nabla^2$), which indicates that $h_{\mu\nu}={\rm diag}[-2\Phi, -2\Psi, -2\Psi, -2\Psi]$ is also a static field, one can obtain the following line element in isotropic coordinates ($t, r, \theta, \phi$)
\begin{equation}
{\rm d}s^{2}=-\left[1+2\Phi(r)\right] {\rm d}t^{2}+\left[1-2 \Psi(r)\right] ({\rm d}r^{2} +r^2{\rm d}\Omega^2), \label{weakmetric}
\end{equation}
where ${\rm d}\Omega^2={\rm d}\theta^{2}+\sin^2\theta \,{\rm d}\phi^{2}$; $\Phi(r)$ and $\Psi(r)$ are radial perturbative potentials satisfying the weak field approximation such that $|\Phi|\ll 1$ and $|\Psi|\ll 1$. Then,  Eq.~\eqref{linfeR}  reduces to  
\begin{equation}
    \nabla^2 h_{\mu\nu}=-16\pi G\, \left(T^{\rm eff}_{\mu\nu}-\frac{1}{2} g_{\mu\nu} T^{\rm eff}\right),
\end{equation}
where $\nabla $ is the three-dimensional gradient operator.
In the Solar system, these potentials have values of about $10^{-5}$ (in geometrized units) at most \cite{Will:2018bme}. Hence, we finally reach
\begin{align}   
\label{wflimFEs}
    \nabla^2 \Phi=8\pi G\, \left(T_{00}^{\rm eff}+\frac{1}{2} T^{\rm eff}\right),   \\
    \nabla^2 \Psi=8\pi G\, \left(T_{11}^{\rm eff}-\frac{1}{2} T^{\rm eff}\right). \label{wflim11}
\end{align}
Furthermore, from Eq.~\eqref{velemt}, the slow motion condition requires
\begin{align}
\label{Tmunus}
T_{00}^{\rm eff} \gg T_{0i}^{\rm eff} \gg T_{ij}^{\rm eff},    
\end{align}
 hence, the components of the effective EMT can be approximately written as $T_{00}^{\rm eff} \simeq \rho_{\rm eff}$, $T_{0i}^{\rm eff} \simeq \rho_{\rm eff} v_i$ and $T_{ij}^{\rm eff} \simeq \rho_{\rm eff} v_i v_j+p_{\rm eff} \delta_{ij}$ (cf. Refs.~\cite{Chandrasekhar:1965gcg,Anderson:1975}). Accordingly, we have $p_{\rm eff} \ll \rho_{\rm eff}$; viz., the condition~\eqref{Tmunus} implies, except $T_{00}^{\rm eff}$, all the other components of $T_{\mu\nu}^{\rm eff}$ are negligible, so that we take $T_{0i}^{\rm eff}=T_{ij}^{\rm eff}=0$ stating that in this approximation, insofar as gravity is concerned, effective momenta, pressure, and stresses are negligible. In line with this, we also have $T^{\rm eff} \approx - T_{00}^{\rm eff}$. Therefore, from Eqs.~\eqref{wflimFEs} and~\eqref{wflim11}, we obtain $\nabla^2 \Phi=\nabla^2 \Psi=4\pi G\, T_{00}^{\rm eff}$ which along with asymptotic flatness implies that $\Psi=\Phi$ in the Newtonian limit. Eventually, the field equations turn into
\begin{equation}
\label{eq:pois}
    \nabla^2 \Phi=4\pi G \rho_{\rm eff},
\end{equation}
which is the Poisson's equation of gravity, and the geodesic equation into
\begin{equation}
\label{eq:new}
    \frac{{\rm d}^2\vec{x}}{{\rm d} t^2}\equiv \vec{a}=-\nabla\Phi,
\end{equation}
which gives the Newtonian acceleration law in a gravitational potential $\Phi$. These two equations reveal an equivalence between Newton's theory and EMPG in the weak field, slow motion limit, but with $\rho_{\rm eff}$ instead of $\rho$.

For a spherically symmetric body surrounded by vacuum---we can safely ignore the cosmologically inferred value of vacuum energy density in the astrophysical setups we are dealing with here---, Eq.~\eqref{eq:pois} leads to
\begin{equation}
\label{eq:Phi}
\Phi=-\frac{G M_{\rm eff}}{r}=-\frac{G }{r}\left(M+ M_{\rm empg}\right),
\end{equation}
where $M_{\rm eff}=4 \pi \int_{0}^{R}\rho_{\rm eff}r^2 {\rm d} r$ and $M=4 \pi \int_{0}^{R}\rho r^2 {\rm d} r$ with $R$ being the radius of the spherical body.\footnote{Here we have dropped the constant potential term as we assume asymptotic flatness  ($\Phi\rightarrow0$ as $r\rightarrow\infty$); also, recall that a constant potential does not affect the acceleration of the test particles.} We would like to note that the vanishing divergence of the effective EMT given in Eq.~\eqref{nonconservedenergy} along with Eq.~\eqref{effemt} guarantees ${\rm d} M_{\rm eff}/{\rm d} t=0$ only in the Newtonian limit. Thereby, in order to assure that $M_{\rm eff}$ does not change with time, the \textit{exact} continuity equation\footnote{Similar to the field equations~\eqref{modfieldeq}, the continuity equation is obtained from the variational principle as well~\cite{Taub:1954, Schutz1970}. Let us express the EoS of the perfect fluid in the form $p=p(h,s)$ where $h=(\rho+p)/\rho_{0}$ is the specific relativistic enthalpy and $\rho_{0}$ is the rest-mass density. The first law of thermodynamics can be written as ${\rm d} p=\rho_0 \,{\rm d} h-\rho_0 \,\mathcal{T}\,{\rm d} s$ where $\mathcal{T}$ is the temperature  and $s$ is the specific entropy. We introduce the Taub vector as $V^{\mu}= h u^{\mu}$, which is defined by five scalar velocity-potential fields ($\phi, \alpha, \beta, \theta, s$) that are independent of the metric tensor~\cite{Schutz1970}. Namely, in the velocity-potential representation, the Taub vector is expressed as $V_{\mu}=\partial_{\mu} \phi+\alpha \, \partial_{\mu} \beta+\theta \, \partial_{\mu} s$, and from the normalization condition of four-velocity, we obtain $h^2=- g^{\mu\nu} V_{\mu} V_{\nu}$. In the presence of perfect fluid described by $\mathcal{L}_{\rm m}=p$, we vary the action~\eqref{action} with respect to the dynamical variable $\phi$ as follows;
%\delta S&=\int {\rm d}^4x\, \frac{\delta{[\sqrt{-g}\, (p+f)]} }{\delta \phi} \delta \phi, \\
\begin{align}
\delta S&&=\int {\rm d}^4x\, \left\{\frac{\partial{[\sqrt{-g}\, (p+f)]} }{\partial \phi}+\partial_{\mu}\left[\frac{\partial{[\sqrt{-g}\, (p+f)]} }{\partial (\partial_{\mu}\phi)}\right] \right\} \delta \phi, \nonumber
\end{align}
where the first term vanishes since the Taub vector depends on the derivative of $\phi$ but not $\phi$ itself. Consequently, substituting $\frac{\partial p}{\partial h}\big|_s=\rho_0$ from the first law of thermodynamics yields
\begin{align}
\delta S=\int {\rm d}^4x\, \left\{\partial_{\mu}\left[\sqrt{-g} \rho_0 \frac{\partial{h} }{\partial (\partial_{\mu}\phi)}\right]+\partial_{\mu}\left[\frac{\partial{(\sqrt{-g}\, f)} }{\partial (\partial_{\mu}\phi)}\right]\right\}  \delta \phi. \nonumber
\end{align} 
Using the definition of $h$ along with the fact that $\nabla_{\mu} V^{\mu}=\frac{1}{\sqrt{-g}} \partial_{\mu}(\sqrt{-g} V^{\mu})$, without any further calculation, it can be seen that $\nabla_{\mu}(\rho_0 u^{\mu}) \neq 0$ in EMSG, instead we have $\nabla_{\mu}(\rho_0 u^{\mu})= - \textnormal{ (terms arisen due to} \,f)$. On the other hand, according to our effective source interpretation, the effective matter Lagrangian density is defined as $\mathcal{L}_{\rm m}^{\rm eff} \equiv p+f(T_{\mu\nu} T^{\mu\nu})=p_{\rm eff}$ and following the same variational procedure with effective variables leads to $\nabla_{\mu}(\rho_{\rm eff0} u^{\mu})=0$.}, which is independent of the conservation of the effective EMT, should also be introduced in EMPG as $\nabla_{\mu}(\rho_{\rm eff0} u^{\mu})=0$ instead of the standard GR form $\nabla_{\mu}(\rho_{0} u^{\mu})=0$ where $\rho_{\rm eff0}$ is the effective rest-mass density defined as $\rho_{\rm eff}=\rho_{\rm eff0}(1+\Pi_{\rm eff})$ with $\Pi_{\rm eff}$ being the specific internal energy---in the Newtonian limit, we drop the subscript 0 since $\rho_{\rm eff} \approx \rho_{\rm eff0}$ when we consider the slow motion approximation $v^2 \sim \Pi_{\rm eff}\sim G M_{\rm eff} / r \sim p_{\rm eff}/\rho_{\rm eff}$. These conditions are satisfied in the Solar System; as is well known, in the Solar System, the effects of the gravitational field are weak, objects (except photons) move slowly and the Sun can be considered to be surrounded by vacuum. So, in the EMPG model applied to the Solar System, $\Phi$ as given by~\eqref{eq:Phi} can be taken as the conventional Newtonian potential when $M_{\rm eff} $ is identified as the ``Newtonian mass" that would be measured via astrophysical (Keplerian) methods such as period--radius relationships for orbits of planets or satellites, augmented by measurements of deflection of light, time delay etc.---though, we remind that Eq.~\eqref{eq:new} does not really apply to photons (they are fast, not slow), so the magnitude of the deflection of a light beam in gravitational field is predicted to be twice as large in GR than in Newton's theory.

In a recent paper \cite{Nazari:2022fbn}, the authors calculate the deflection of light by considering $M=4 \pi \int_{0}^{R}\rho r^2 {\rm d} r$ as the mass measured from astrophysics in order to recover the GR value of the deflection in case of $\alpha=0$, such that EMSG modification $M_{\rm empg}$ is attached to the $\gamma$ parameter resulting in its deviation from unity. However, it is clear from the discussions in the above paragraphs that if the observer does not know the energy density or the mass of the astrophysical object in an independent way except for observation of orbits or similar measurements, the observer cannot distinguish whether the mass of the astrophysical object is $M_{\rm eff}$ or $M$, i.e., whether the curvature of the spacetime is governed by GR or EMPG. To prevent such a misinterpretation of the PPN parameter $\gamma$, this proper reasoning should be followed both in any particular model of EMSG theory and in models of matter--curvature coupling gravity, if matter is minimally coupled to curvature, in other words, if the modification to GR is of the form $f(\mathcal{L}_{\rm m},T,T^{\mu\nu}T_{\mu\nu})$, i.e., is a function of only matter-related terms.

Next, substituting $\rho_{\rm eff}$ defined in Eq.~\eqref{eq:effdens} into Eq.~\eqref{eq:pois}, we can also write the Poisson's equation in terms of the standard matter variables:
 \begin{equation}
\begin{aligned}
\label{wflimofempgs}
    \nabla^2 \Phi=4\pi G\bigg\{\rho-\alpha (\rho^2+3p^2)^{\eta}\bigg[1-2\eta \bigg(1+\frac{4 \rho p}{\rho^2+3p^2}\bigg)\bigg]\bigg\}.
    \end{aligned}
\end{equation}
As we have $T_{ii}^{\rm eff}=0$, but not $T_{ii}=0$ in the Newtonian limit, we see not only $\rho$, but also $p$ in Eq.~\eqref{wflimofempgs}, in contrast to GR ($\alpha=0$). We remark that Ref.~\cite{Kazemi:2020hep} briefly discusses the weak field limit of the quadratic EMSG model (viz., $\eta=1$ case of the EMPG). In that study, however, the authors limit the scope of the discussion by assuming $|p/\rho|\ll1$ from the outset. The full scope of the discussion in the framework of the EMPG model requires the use of a more general condition than $|p/\rho|\ll1$, that is,
    \begin{align}
    \label{eqn:concon}
    \bigg|\frac{p_{\rm eff}}{\rho_{\rm eff}}\bigg|\ll1,
    \end{align}
 and we employ the latter in this work. The condition given in Eq.~\eqref{eqn:concon} does not necessarily imply that $|p/\rho|\ll1$; except of course, in the GR limit ($\alpha\rightarrow0$) of the EMPG model. Yet, it is still useful and would be realistic to describe most of well known astrophysical objects such as planets, stars possessing negligible pressure in comparison with their energy densities. In the Solar System, within the context of GR, typical $p/\rho$ values are $10^{-10}$ for the Earth, $10^{-5}$ for the Sun. Moreover, this ratio for the surface of a white dwarf is about $10^{-4}$~\cite{Poisson:2014book}. It is, on the other hand, $\sim 0.1$ for the surface of a neutron star\footnote{See Figure 1(b) in Ref.~\cite{Akarsu:2018zxl} for effective EoS versus the energy density of the neutron star matter stress where the case $\alpha=0$ (GR) gives the EoS of the matter stress itself for various realistic EoS parametrizations. The surface value of $p/\rho$ is achieved for $p=0$. For $\alpha\neq0$ in quadratic EMSG model, EoS parameter on the surface does not deviate from GR, while deviations are seen on the core values in some parametrizations.} and $\sim 1$ for the event horizon of a blackhole based on Zeldovich (stiff) fluid ($p=\rho$) which is the most rigid EoS compatible with the requirements of relativity~\cite{zeldovich}. We proceed with the general case which provides us with a mathematical simplicity for applying our results to the astrophysical objects with structures and postpone the discussion of the limit $|p/\rho|\ll1$ until Section \ref{discussion}.
 
\subsubsection{Post-Newtonian limit}
The Newtonian limit  of any gravity theory should be able to explain aspects of light propagation such as deflection of light and Shapiro time delay in addition to the Newtonian properties of planetary and satellite orbits. However, for cases like the perihelion shift of Mercury in which the more accuracy is needed, the linearized theory is inadequate and one must include the second order perturbation terms as well.

Accordingly, let us proceed with writing the Ricci tensor in the second order $h_{\mu\nu}$'s as~\cite{Carroll:1997ar}
\begin{align}
\mathcal{R}^{(2)}_{\mu\nu}&=\frac{1}{2}\bigg\{ h^{\alpha\beta}  (\partial_{\mu} \partial_{\nu} h_{\alpha\beta}-  \partial_{\alpha} \partial_{\mu} h_{\nu\beta}-\partial_{\alpha} \partial_{\nu} h_{\mu\beta})  \nonumber \\
&+\frac{1}{2} \partial_{\mu} h_{\alpha\beta} \partial_{\nu} h^{\alpha\beta}+\partial^{\alpha} h_{\beta\nu} (\partial_{\alpha} h^{\beta}_{\mu}- \partial^{\beta} h_{\alpha\mu}) \nonumber  \\
&+\left[\partial_{\alpha} h^{\alpha\beta}-\frac{1}{2}\eta^{\alpha\sigma}\partial^{\beta} h_{\alpha\sigma}\right] (\partial_{\beta} h_{\mu\nu}-\partial_{\mu} h_{\nu\beta}-\partial_{\nu} h_{\mu\beta})\bigg\}.
\end{align}
Including the second order perturbations in Lorenz gauge implies $h^{\mu\nu}(\partial_{\mu} h_{\alpha\nu}+\partial_{\nu} h_{\alpha\mu}-\partial_{\alpha} h_{\mu\nu})=0$, and hence, the second order Ricci tensor becomes
\begin{align}
\mathcal{R}^{(2)}_{\mu\nu}=&\frac{1}{2} h^{\alpha\beta}  \partial_{\mu} \partial_{\nu} h_{\alpha\beta}
+\frac{1}{4} \partial_{\mu} h_{\alpha\beta} \partial_{\nu} h^{\alpha\beta}  \nonumber  \\
&+\frac{1}{2} \partial^{\alpha} h_{\beta\nu} (\partial_{\alpha} h^{\beta}_{\mu}- \partial^{\beta} h_{\alpha\mu}).
\end{align}
Then, the $tt$ component of Eq.~\eqref{FEwithR} up to second order perturbations reads
\begin{align}
\nabla^2 \Phi^{(2)}-2 \left(\frac{{\rm d} \Phi}{{\rm d} r}\right)^2=4 \pi G \rho_{\rm eff},  % \sim  O(\epsilon^2) , 
\end{align}
and yields the solution 
\begin{align}
\Phi^{(2)}=-\frac{ G M_{\rm eff}}{r}+\frac{ G^2 M_{\rm eff}^2}{r^2} .  \end{align}
Therefore, in the EMPG model, the components of the post-Newtonian metric take the following form:
\begin{align}
g_{00}&=-1+\frac{2 G M_{\rm eff}}{r}- \frac{2 G^2 M_{\rm eff}^2}{r^2} + \mathcal{O}\left(\frac{1}{r^3}\right) , \label{g00} \\
g_{11}&=1+\frac{2 G M_{\rm eff}}{r} + \mathcal{O}\left(\frac{1}{r^2}\right). \label{g11}
\end{align}

We note that with this solution, the line element~\eqref{weakmetric} approaches the Minkowski spacetime metric as $r \rightarrow \infty$ and has exactly the same $r$ dependence as in GR. Conversely, this case does not apply to gravity models with curvature-type modifications. For instance, $f(\mathcal{R})$ type models give rise to an extra term of the form $\propto e^{-m r}/r$ with $m$ being a constant, like Yukawa type potentials, and therefore change the $r$ dependency of the gravitational potential and require two different parameters, instead of the single $M_{\rm eff}$ parameter in EMPG, to fix the solution~\cite{Sbisa:2019mae}.

\subsubsection{Parametrized Post-Newtonian formalism} \label{sec:ppn}
Through the PPN formalism, gravitational theories can be distinguished from each other by the numerical coefficients appearing in front of the metric potentials. As first formulated by Eddington, Robertson and Schiff \cite{Eddington24,Robertson62,Schiff67}, and later fixed systematically by the approach of Nordvedt and Will \cite{Nordvedt68,Will71,Will72} in this formalism, the coefficients are replaced by various compositions of a total of ten parameters whose values depend on the gravity theory under consideration. They used it in interpreting the Solar System experiments. The PPN metric up to the order we have calculated above is written as follows
\begin{align}
g_{00}&=-1+2U-2 \beta U^2,  \\
g_{11}&=1+2 \gamma U.
\end{align}
As is seen from Eqs.~\eqref{g00} and~\eqref{g11}, in EMPG, we have 
\begin{equation}
U=\frac{G M_{\rm eff}}{r},
\end{equation}
which implies that
\begin{equation}
\gamma=1\quad\textnormal{and}\quad    \beta=1,
\end{equation}
exactly the same values in GR which in turn are consistent with the experimental and
observational bounds \cite{Will:2014kxa,Will:2018bme}.

The PPN parameter $\gamma$, which is the measure of space-curvature produced by unit rest mass, plays a crucial role in the Solar System tests of a gravitational theory. In GR, we know that a light ray passing near the Sun is deflected by some angle (deflection of light) and a radar signal from Earth sent on a round-trip passing near the Sun requires some extra time compared to the time interval predicted from Newtonian theory (time delay of light). Both these deflection angle and delayed time is proportional to the coefficient $\frac{1}{2} (1+\gamma)$. The factor $\frac{1}{2}$ here appears in any metric theory of gravity while the factor $\frac{\gamma}{2}$ changes from one theory to the other. The other parameter $\beta$, which is the measure of nonlinearity in the superposition law for gravity, is related to the perihelion shift of Mercury through the coefficient $\frac{1}{3} [2(1+\gamma)- \beta]$. Due to the observational limits, other PPN parameters should be very close to zero in alternative theories of gravity, and they are automatically zero in EMPG with nonnegative $\eta$, as in GR.

\section{Schwarzschild Exterior Solution} 
\label{sec:exterior}

The particular models of EMSG satisfying $f(T_{\mu\nu} T^{\mu\nu})=0$ when $T_{\mu\nu}=0$ are equivalent to GR in vacuum. Obviously, this is also the case for the EMPG model with $\eta\geq0$. Therefore, assuming $\eta\geq0$, the field equations of the EMPG model for vacuum satisfy the spherically symmetric and static Schwarzschild line element, in Schwarzschild coordinates ($t, \bar{r}, \theta, \phi$),
\begin{align}
{\rm d}s^{2} = -\left(1-\frac{2 G M_{\rm eff}}{\bar{r}}\right) {\rm d}t^{2} + \frac{1}{\left(1-\frac{2 G M_{\rm eff}}{\bar{r}}\right)} {\rm d}\bar{r}^{2} + \bar{r}^2 {\rm d}\Omega^2,  \label{vacsoln}
\end{align}
where $M_{\rm eff}$ is an integration constant that is subject to be determined from astronomical/astrophysical measurements, e.g., the period and semi-major axis of a planet's orbit. $M_{\rm eff}$ also corresponds to the volume integral of $\rho_{\rm eff}$ by means of the $tt$ component of the field equations~\eqref{fieldeq}, viz., $M_{\rm eff}=4 \pi \int_{0}^{R}\rho_{\rm eff}r^2 {\rm d} r$ for a spherical object.  This vacuum solution can also be written in isotropic coordinates via the transformation $\bar{r}=r \left(1+\frac{G M_{\rm eff}}{2 r}\right)^2$~\cite{Poisson:2014book}, and we can obtain the Newtonian and post-Newtonian limits of EMPG in this way as well. In isotropic coordinates, the expression given in Eq.~\eqref{vacsoln} turns out to be
\begin{equation}
\begin{aligned}  \label{metriciso}
ds^{2} =&-\left(\frac{1-\frac{G M_{\rm eff}}{2 r}}{1+\frac{G M_{\rm eff}}{2 r}}\right)^2 {\rm d} t^{2}+ \left(1+\frac{G M_{\rm eff}}{2 r}\right)^4 ({\rm d} r^{2} +r^2 {\rm d}\Omega^2). 
\end{aligned}
\end{equation}
In the weak field region, that is, at distances far away from the massive object which corresponds to $r \gg G M_{\rm eff}$, the metric encoded in Eq.~\eqref{metriciso} can be expanded as
\begin{align} \label{Sch00}
&g_{00}=-1+  \frac{2G M_{\rm eff}}{r}  -\frac{2G^2 M_{\rm eff}^2}{r^2}+\mathcal{O}\left(\frac{1}{r^3}\right), \\
&g_{11}=1+  \frac{2G M_{\rm eff}}{r}+\mathcal{O}\left(\frac{1}{r^2}\right). \label{Sch11}
\end{align}

We see that the metric components given in Eqs.~\eqref{Sch00} and~\eqref{Sch11} are identical to the ones given in Eqs.~\eqref{g00} and~\eqref{g11}, hence, give rise to the same PPN parameters obtained from weak field, slow motion limit (see Section~\ref{sec:ppn}). This is a result in line with our claim that the slow motion condition should be written in terms of effective variables. It is also worth noting here that the only difference between the line element---expressed in either Schwarzschild coordinates~\eqref{vacsoln} or isotropic coordinates~\eqref{metriciso}---, and the one obtained in GR is that where the term $M_{\rm eff}$ is, there is $M$. Both are not directly observable parameters but are parameters derived using EMPG or GR from observational data on Kepler orbits, deflection of light etc. That is, using the same astronomical data will result in exactly the same mass value whether EMPG is used or GR is used, but this mass value will be the value of $M_{\rm eff}$ in EMPG and the value of $M$ in GR. In the case of EMPG, if there is information about the values of the EMPG parameters $\alpha$ and $\eta$ from another independent physical event (e.g., from cosmological observations), then it is possible to separate the value of $M_{\rm eff}=M+M_{\rm empg}$ into the values of its components $M$ and $M_{\rm empg}$.

\section{Slow motion conditions}
\label{discussion}
 Assuming a barotropic fluid, viz., a fluid with an EoS of the form $p=p(\rho)$, and $p(\rho=0)=0$, we have already shown that $\eta$,  the power of the Lorentz scalar $T_{\mu\nu}T^{\mu\nu}$, is necessarily nonnegative for $\rho_{\rm eff}$ and $p_{\rm eff}$ not to be divergent [see Eq.~\eqref{eq:effdens}] as $\rho\rightarrow0$. Therefore, we must have $\eta \geq 0$ to be able to obtain a viable effective source within EMPG. In order to determine the slow motion condition of EMPG, we have used the effective energy density and pressure of the massive body since these are the values that are measurable in the long-range regime. We now analyse and discuss effects of EMPG corrections by considering the standard $\rho$ and $p$ accompanied by parameters $\alpha$ and $\eta$ in contrast to GR ($\alpha=0$). Besides the slow motion condition $|p_{\rm eff}/\rho_{\rm eff}|\sim v^2 \ll 1$ [see Eq.~\eqref{velemt}], we also have to assign a relationship between $p$ and $\rho$ specifying the EoS $p=p(\rho)$. Let us proceed by writing the effective energy density and pressure \eqref{eq:effdens} in the following form: 
\begin{align}
\rho_{\rm eff}&=\rho+\rho_{\rm empg}, \label{genrhoeff} \\
p_{\rm eff}&=p+\frac{\rho_{\rm empg}}{2\eta \left(1+\frac{4 p/\rho}{1+3p^2/\rho^2}\right)-1} \label{genPeff}, 
\end{align}
 where $\rho_{\rm empg}$ represents the extra terms arising from the EMPG modification, viz.,
\begin{equation}
\begin{aligned}  
\label{genrhoempg}
\rho_{\rm empg}=&\,\alpha'\bar{\rho}\left(\frac{\rho}{\bar{\rho}}\right)^{2\eta}  \left(1+\frac{3 p^2}{\rho^2}\right)^{\eta}\\&\times\bigg[2\eta\bigg(1+\frac{4 p/\rho}{1+3p^2/\rho^2}\bigg)-1\bigg].
\end{aligned}
\end{equation}
We should point out that although the last term in Eq.~\eqref{genPeff} with a possibly vanishing denominator seems problematic at first glance, substituting Eq.~\eqref{genrhoempg} back into Eq.~\eqref{genPeff} drops this factor. Here the dimensionless coupling parameter is defined as 
\begin{align}
 \alpha'=\alpha\bar{\rho}^{\;2\eta-1},
\end{align}
where $\bar{\rho}$ may be taken as, e.g., $\bar{\rho}=10^3 \,{\rm kg/m^3}$, referring the mass (since $c=1$) of 1 $\rm m^3$ water, a scale consistent with the average densities of the Solar System bodies, such as the Sun and planets---so that we use a scale that can be defined with respect to an object's mass which is known/measured by a method that does not include gravitational effects. Note that the dimensionality of the parameter $\alpha$ depends on $\eta$ which renders it unreasonable to make a comparison between magnitude of $\alpha$ for different models; through the redefinition of the model parameter above, we overcome this issue. Next, we eliminate $\rho_{\rm empg}$ in Eqs.~\eqref{genrhoeff} and~\eqref{genPeff}, hence, realize that independent of $\alpha'$, $\rho_{\rm eff}$ and $\rho$ are related as follows:
\begin{align} \label{genrhorhoeff}
\rho_{\rm eff} &\left\{1+ \left[1-2\eta \left(1+\frac{4 p/\rho}{1+3p^2/\rho^2}\right)\right] \frac{p_{\rm eff}}{\rho_{\rm eff}}\right\} \nonumber   \\
&=\rho \left\{1+  \left[1-2\eta \left(1+\frac{4 p/\rho}{1+3p^2/\rho^2}\right)\right] \frac{p}{\rho}\right\}.
\end{align}

We shall henceforth focus on a particular class of EoS's satisfying the condition
 \begin{align}
 \label{astrcond}
    \left|\frac{p}{\rho}\right|\ll 1.
\end{align} 
If one considers that the pressure inside the nonrelativistic astrophysical objects is much less than the corresponding energy density, this is indeed a rather natural assumption. Therefore, using Eq.~\eqref{astrcond} in the effective energy density~\eqref{genrhoeff} and pressure~\eqref{genPeff}, these become
 \begin{align}
\rho_{\rm eff}&=\rho+\rho_{\rm empg}, \label{effrho00}  \\
p_{\rm eff}&=p+\frac{\rho_{\rm empg}}{2\eta-1}, \label{effP0}
\end{align}
where the energy density arising from the EMPG modification~\eqref{genrhoempg} reads
\begin{align}
\label{eq:rhoempg}
\rho_{\rm empg}=(2\eta-1)\alpha'\bar{\rho}\left(\frac{\rho}{\bar{\rho}}\right)^{2\eta}.
\end{align}
As mentioned above, Eq.~\eqref{effP0} seems to explode for $\eta=1/2$, but that is not actually the case; substituting Eq.~\eqref{eq:rhoempg} into Eq.~\eqref{effP0}, it can be seen immediately that the factor $2\eta-1$ responsible for this situation drops. Accordingly, the slow motion condition can be restated for a general $\eta$ in terms of the standard energy density and pressure as
\begin{equation}
\label{effw0}
    \left|\frac{p_{\rm eff}}{\rho_{\rm eff}}\right|=\left|\frac{p+(2\eta-1)^{-1}\rho_{\rm empg}}{\rho+\rho_{\rm empg}} \right|\ll 1,
\end{equation}
from which, however, it is not easy to see the validity of slow motion condition. Instead, manipulating Eqs.~\eqref{effrho00} and~\eqref{effP0}, or equivalently setting $|p/\rho| \ll 1$ in Eq.~\eqref{genrhorhoeff}, we obtain
\begin{align}
\label{rhorhoeff}
\rho_{\rm eff} \left[1+ (1-2\eta) \frac{p_{\rm eff}}{\rho_{\rm eff}}\right]=\rho \left[1+  (1-2\eta) \frac{p}{\rho}\right],
\end{align}
which turns out to be quite useful. As can be seen from Eq.~\eqref{rhorhoeff}, if $p_{\rm eff}/\rho_{\rm eff}\approx p/\rho$, then $\rho_{\rm eff} \approx \rho$ independent of the value of $\eta$. Otherwise, $\rho_{\rm eff}$--$\rho$ relation depends on $\eta$ and $\rho_{\rm eff}$ can even take negative values. Having said that, we know from cosmological analyses to date that $\eta$ is expected to be at the order of $\mathcal{O}(\eta)=1$. In Ref.~\cite{Akarsu:2017ohj}, the authors show that $\eta\sim0$ can explain the late-time acceleration of the universe from the dust's EMPG contribution (in this case, since it resembles the cosmological constant), without resorting to a cosmological constant ($\Lambda$) or dark energy and find that $\eta=-0.003\pm0.023$ at 95\% confidence level from their observational analyses. In a more recent cosmological analysis~\cite{Kolonia:2022jje}, it is obtained $0<\eta<0.18$ at 95\% confidence level, and $\eta=0.26\pm0.25$ at 95\% confidence level when a $\Lambda$ is allowed on top of EMPG. In the case $\eta=1$, in the modified Friedmann equation, the dust energy density $\rho_{\rm m}\propto a^{-3}$ is accompanied by the additional energy density term mimicking stiff (Zeldovich) fluid $\rho_{\rm emsg}\propto a^{-6}$ (see the higher-order correction terms on effective EMT given in Eq.~\eqref{rhopeff} which are proportional to $\rho^{2\eta}$ for arbitrary $\eta$). The present-day density parameter of such stiff fluid-like sources (effective or actual), $\Omega_{\rm stiff}$, on top of the standard $\Lambda$CDM model are extremely well constrained; assuming it is non-negative definite, the upper bounds on it ranges from $\sim10^{-4}$ to $\sim10^{-18}$ from cosmological analysis depending on the datasets used (e.g., SnIa, BAO, Planck CMB); and even reaches $\sim10^{-23}$ when big bang nucleosynthesis is considered~\cite{Akarsu:2019pwn,Akarsu:2021max,ABLK23}. Similarly, the constraints on the EMPG with $\eta=1$ from neutron stars suggest that the new terms arising in this case must be extremely weakly coupled with gravity, viz., $|\alpha|\lesssim 10^{-19} {\rm m^3kg^{-1}}$. From all this, it would be fair to say that $\eta > 1$ cases are unlikely to be realistic. Therefore, in the light of the above discussion on theoretical and observational bases, we do not expect EMPG to be realistic out of interval of $0 \lesssim \eta \lesssim 1$. Consequently, if we also consider the $\eta\geq0$ condition we introduced for the non-divergent $\rho_{\rm eff}$ and $p_{\rm eff}$ in vacuum solutions, then the interval $0\leq\eta\lesssim1$ should be taken for realistic implementations of the EMPG model. Now, with this condition in mind, we can continue our investigation in this chapter.

When we consider the conditions $|p_{\rm eff}/\rho_{\rm eff}|\ll 1$ (implied by the slow motion condition) and $|p/\rho| \ll 1$ (expected to be applied to most astrophysical objects~\cite{Will:2018bme}; however, exceptions are possible where it will not apply under extreme conditions, such as the deep interior of neutron stars, see, e.g., \cite{Akarsu:2018zxl}) simultaneously, Eq.~\eqref{rhorhoeff} implies that
\begin{equation}
 \rho_{\rm eff} \approx \rho,
\end{equation}
for $\mathcal{O}(\eta)=1$ (required for the realistic implementations of EMPG). Then we see that, if we only demand that the slow motion condition (implying $|p_{\rm eff}/\rho_{\rm eff}|\ll 1$) to be satisfied, to get $\rho_{\rm eff} \approx \rho$ from  Eq.~\eqref{effrho00}, $\rho_{\rm empg}$ must remain negligible compared to $\rho>0$, viz., we must have
\begin{align}   \label{rho-sm}
 |\rho_{\rm empg}|\ll \rho,
\end{align}
which also guarantees that $\rho_{\rm eff}>0$ as $\rho$ is positive definite by definition. Using these findings, we can also write some conditions between $\alpha'$ and $\eta$:

\textbf{(i)} The case $0<\eta<1/2$: When we use Eq.~\eqref{rho-sm} in Eq.~\eqref{eq:rhoempg}, we can write
\begin{align}
\label{eq:c1}
  |\alpha'| \ll \frac{(\rho/\bar{\rho})^{1-2\eta}}{1-2\eta}.
\end{align}
From this inequality we see that the lower energy density, the smaller $|\alpha'|$ must be. This is more emphasized for smaller $\eta$ values.

\textbf{(ii)} The case of Scale-Independent EMSG (viz., $\eta=1/2$): This requires a separate examination, as we cannot use Eq.~\eqref{rhorhoeff} in this case. To do so, we first substitute Eq.~\eqref{eq:rhoempg} in Eqs.~\eqref{effrho00} and~\eqref{effP0}, and then choose $\eta=1/2$ to get the relation we need;
\begin{equation}  
     \left|\frac{p_{\rm eff}}{\rho_{\rm eff}}\right|= \left|\frac{p}{\rho}+\alpha'\right| \ll 1.
\end{equation}
This in turn implies that we must have
\begin{equation}
\label{eq:c2}
    |\alpha'|\ll1,
\end{equation}
regardless of the energy density scale considered.

\textbf{(iii)} The case $1/2<\eta \lesssim 1$: When we use Eq.~\eqref{rho-sm} in Eq.~\eqref{eq:rhoempg}, we can write
\begin{align}
\label{alpha3}
  |\alpha'| \ll -\frac{(\rho/\bar{\rho})^{1-2\eta}}{1-2\eta},
\end{align}
which differs from Eq.~\eqref{eq:c1} with a minus sign. From this inequality we see that the higher energy density, the smaller $|\alpha'|$ must be. This is more emphasized for larger $\eta$ values.

\section{Conclusion}   
\label{conclusion}

We have explored the weak field and slow motion limits,  Newtonian and Post-Newtonian limits, of the EMPG \cite{Akarsu:2017ohj,Board:2017ign}, namely, the EMSG \cite{Katirci:2014sti,Akarsu:2017ohj,Board:2017ign} of the form $f(T_{\mu\nu}T^{\mu\nu})=\alpha (T_{\mu\nu}T^{\mu\nu})^{\eta}$, where $\alpha$ (determines the gravitational coupling strength of the EMPG modification) and $\eta$ are constants. In PPN formalism, we have shown that EMPG with $\eta\geq0$ (otherwise the EMPG modification would diverge in vacuum solutions) and GR are not distinguishable by local tests, say, the Solar System tests; as they lead to \textbf{(i)} the same form of gravitational potential, keeping PPN parameters completely the same in both theories, and \textbf{(ii)} the same geodesics for the test particles. However, within the EMPG framework, the mass of an astrophysical object inferred from astronomical observations $M_{\rm ast}$, e.g., using Keplerian methods, corresponds to the effective mass $M_{\rm eff}(\alpha,\eta,M)$ (viz., the mass of the object resulting from the \textit{effective} EMT describing it, $T_{\mu\nu}^{\rm eff}$) with $\alpha$ and $\eta$ being the free parameters of EMPG and $M$ being the actual \textit{physical} mass (viz., the mass of the object resulting from the \textit{actual} EMT describing it, $T_{\mu\nu}$). Accordingly, while in the GR framework we simply have the relation $M_{\rm ast}=M$, in the EMPG we have $M_{\rm ast}=M+M_{\rm empg}$. In the case of EMPG, if there is information about the values of $\{\alpha,\eta\}$ pair or $M$ from other independent phenomena (from cosmological observations, structure of the astrophysical object, etc.), then in principle it is possible to infer not only $M_{\rm ast}$ alone from astronomical observations, but also $M$ and $M_{\rm empg}$ separately. We have concluded also that for a proper analysis within EMPG framework, which leads us to the results we have stated above, it is necessary to describe the slow motion condition, which is also related to the Newtonian limit approximation, by $|p_{\rm eff}/\rho_{\rm eff}|\ll1$, whereas it is by $|p/\rho|\ll1$ in GR---note that the latter need not be satisfied for the former to be satisfied.

Finally, although the current work is based on a specific model of EMSG theory \cite{Katirci:2014sti,Akarsu:2017ohj,Board:2017ign}, namely the EMPG model \cite{Akarsu:2017ohj,Board:2017ign}, it is conceivable that our findings would apply to models of EMSG in general---and further to modified gravity theories such as $\mathcal{R}+f(\mathcal{L}_{\rm m})$ \cite{Harko:2010mv} and $\mathcal{R}+f(T)$ \cite{Harko:2011kv} as all these are similar in that the source is minimally coupled to the curvature. Consequently, apart from several specific conclusions we have drawn from this study, we have learned two important lessons; in studies aimed at constraining the free parameters of such modified theories of gravity, using astronomical information, caution must be exercised in drawing conclusions, and then such theories may enjoy a significant advantage, as they escape local tests.

\begin{acknowledgments}
Valuable comments and suggestions by the referees are gratefully acknowledged. The authors thank Elham Nazari for useful discussions. \"{O}.A. acknowledges the support by the Turkish Academy of Sciences in the scheme of the Outstanding Young Scientist Award  (T\"{U}BA-GEB\.{I}P). N.K. thanks Do\u gu\c s University for the financial support provided by the Scientific Research (BAP) project number 2021-22-D1-B01. \"O.A. and N.K. are supported in part by T\"UB\.ITAK grant 122F124. N.M.U. is supported by Boğazi\c ci University Research Fund Grant Number 18541P. \"{O}.A. and N.K. acknowledge the COST Action CA21136 (CosmoVerse).
\end{acknowledgments}

\end{document}